# Constructing Hybrid Incremental Compilers for Cross-Module Extensibility with an Internal Build System

Jeff Smits[a], Gabriël Konat[a], and Eelco Visser[a]

a Delft University of Technology, The Netherlands

**Abstract**

**Context** Compilation time is an important factor in the adaptability of a software project. Fast recompilation enables cheap experimentation with changes to a project, as those changes can be tested quickly. Separate and incremental compilation has been a topic of interest for a long time to facilitate fast recompilation.

**Inquiry** Despite the benefits of an incremental compiler, such compilers are usually not the default. This is because incrementalization requires cross-cutting, complicated, and error-prone techniques such as dependency tracking, caching, cache invalidation, and change detection. Especially in compilers for languages with cross-module definitions and integration, correctly and efficiently implementing an incremental compiler can be a challenge. Retrofitting incrementality into a compiler is even harder. We address this problem by developing a compiler design approach that reuses parts of an existing non-incremental compiler to lower the cost of building an incremental compiler. It also gives an intuition into compiling difficult-to-incrementalize language features through staging.

**Approach** We use the compiler design approach presented in this paper to develop an incremental compiler for the Stratego term-rewriting language. This language has a set of features that at first glance look incompatible with incremental compilation. Therefore, we treat Stratego as our critical case to demonstrate the approach on. We show how this approach decomposes the original compiler and has a solution to compile Stratego incrementally. The key idea on which we build our incremental compiler is to *internally* use an incremental build system to wire together the components we extract from the original compiler.

**Knowledge** The resulting compiler is already in use as a replacement of the original whole-program compiler. We find that the incremental build system inside the compiler is a crucial component of our approach. This allows a compiler writer to think in multiple steps of compilation, and combine that into an incremental compiler almost effortlessly. Normally, separate compilation à la C is facilitated by an *external* build system, where the programmer is responsible for managing dependencies between files. We reuse an existing sound and optimal incremental build system, and integrate its dependency tracking *into* the compiler.

**Grounding** The incremental compiler for Stratego is available as an artefact along with this article. We evaluate it on a large Stratego project to test its performance. The benchmark replays edits to the Stratego project from version control. These benchmarks are part of the artefact, packaged as a virtual machine image for easy reproducibility.

**Importance** Although we demonstrate our design approach on the Stratego programming language, we also describe it generally throughout this paper. Many currently used programming languages have a compiler that is much slower than necessary. Our design provides an approach to change this, by reusing an existing compiler and making it incremental within a reasonable amount of time.

## 1 Introduction

Compilation time of a software project is an important factor in how easily the project can be changed. When the compilation time is low, it is cheap to experiment with changes to the project, which can be tested immediately after recompilation. Therefore, fast recompilation has been a topic of interest for a long time [1, p. 384].

Already in the early times of FORTRAN (as of FORTRAN II [1, p. 384]), *independent compilation* of modules was used to speed up recompilation. This allowed a change in one module to require only the recompilation of that module and linking against the previously compiled other modules of the program. Skipping the compilation of all modules except the changed one was a significant improvement over compiling everything again. This did come at the cost of possible link-time issues. To be able to link the program together, the modules needed to be up-to-date with the data layout defined in `COMMON`. This was done manually and only checked by the programmer.

Mesa introduced *separate compilation*, which solved this issue by moving the static checks of cross-module dependencies to compile-time [7]. When a Mesa module is compiled, the result includes a symbol file that can be used during the compilation of other modules that depend on that module. Other languages such as ML and C use interface files that are written by the programmer. Separate compilation brought back type correctness for statically typed programs, while preserving fast recompilation from *independent compilation*.

To further speed up recompilation, we can save intermediate results during compilation. If parts of the program do not change, then the intermediate results of those parts can be reused. The term *separate compilation* applies to compilation where intermediate results are saved per file [7]. For sub-file level tracking of changes and intermediate results, the term *incremental compilation* is used [17, 18].

Incremental compilation has clear benefits for recompilation speed. However, to make a compiler incremental, we must split the compilation process into separate (ideally independent) parts, track dependencies between these parts, cache previous results, perform cache invalidation, persist caches to disk, detect changes, and propagate them. Unfortunately, this is far from trivial, as these techniques are complicated and error-prone, and cross-cut the concerns of the compiler. Especially in a compiler for a language requiring cross-module linking and integration, correctly and efficiently implementing an incremental compiler can be challenging.

In this paper, we present an approach to the design of incremental compilers that separates the language-specific aspects of a compilation schema from the language-independent aspects of tracking the impact of a change on the output of compilation. Our compilation method is a hybrid of separate and incremental compilation. It is separate compilation in the sense that we process a changed file in its entirety. However, instead of producing a single intermediate representation for a file, we split the intermediate representation into smaller units (e.g. top-level definitions) and separate summaries for static analysis. By splitting up the file early, we can then proceed to process each unit separately. Changing one of the units within a file will only require the reading and splitting up of that file, after which further processing can proceed incrementally. We use PIE [14, 15] – an efficient, precise, and expressive incremental build system – to tie together the components of the compiler in order to efficiently propagate changes through the compiler pipeline. We have developed this approach to incrementalize the compiler of Stratego [2, 3], a term rewriting language with open extensibility features. Our approach allows us to reuse almost all of the existing compiler while gaining great improvements in recompilation speed. We evaluate the incremental Stratego compiler with a benchmark on the version control history of a large Stratego project.

In summary, the paper provides the following contributions:

- We present a general design approach for constructing a hybrid separate/incremental compiler from compiler components and a build system.
- We present an application of the approach in the design and implementation of a hybrid separate/incremental compiler for the Stratego transformation language, for which previously only a whole program compiler was available.
- We evaluate the performance of the Stratego incremental compiler applied to a version history of the WebDSL code base. While the from-scratch time of the incremental compiler is longer than the original compiler, the recompilation time of the incremental compiler is less than 10 % of the from-scratch time in most cases.

**Outline** We proceed as follows. In the next section we discuss the open extensibility features of the Stratego languages, their application to modular language definition, and indicate how these features affect separate/incremental compilation. In section 3, we analyse the interaction between compilers and build systems in general, and discuss the whole program Stratego compiler and a previous attempt at making it incremental. In section 4, we describe our incremental compilation design approach and its application to Stratego. In section 5, we compare the performance of the incremental compiler to the performance of the original, whole-program Stratego compiler, and demonstrate the effectiveness of our solution for successive compilations. In section 6, we discuss related work.

## 2 Open Extensibility in Stratego

The goal of the *expression problem* as formulated by Wadler [26] “is to define a datatype by cases, where one can add new cases to the datatype and new functions over the datatype, without recompiling existing code, and while retaining static type safety (e.g., no casts).” The example used to illustrate the expression problem is the modular definition of a language consisting of a data type for its abstract syntax and operations such as evaluation and pretty-printing on that data type. The problem illustrates the difference in capabilities between object-oriented programming, in which data extension is easy, and functional programming, in which extension with operations is easy.

The Stratego transformation language was designed to support modular language definition with open extensibility. A module can extend a language definition with new AST constructors and/or with new operations, as we will illustrate below. Alas, the language design does not count as a solution to the expression problem since it requires whole program compilation and the language is dynamically typed. In this paper, we address the problem of incrementally compiling Stratego programs in the face of cross-module extensibility of operations (aka strategies), bringing it closer to a solution of the expression problem. In this section, we illustrate the use of open extensibility in Stratego, we discuss the language features that enable that and how they affect separate/incremental compilation, and we discuss a real world application of open extensibility in the WebDSL compiler.

**An Example** We illustrate Stratego's extensibility with a fragment from the TFA example language borrowed from Visser [24]. The language definition in figure 1 is organized in a matrix where each cell corresponds to a Stratego module. The columns correspond to language aspects and the rows define data types or transformations. The modules in the top row define the abstract syntax of a language aspect: a core language with variables and function calls, arithmetic expressions, and control-flow constructs.[1] The modules in the second row define the desugar transformation. Arithmetic operator applications are desugared to function applications (taking the AST constructor as function name, using the pattern `f#(ts)` that generically deconstructs a term into its constructor `f` and child terms `ts`), for loops are desugared to while loops with the appropriate initialization, and if-then statements are desugared to if-then-else statements. The modules in the third row define the eval transformation using rewrite rules for basic reductions and a strategy to define evaluation order. (The details are not relevant for the topic of the paper.) The TFA example described by Visser [24] provides more (interesting) language extensions and operations, but figure 1 demonstrates the essential extensibility features.

The example illustrates how we can orthogonally extend a language definition with new constructors and/or transformations. We discuss the language features that enable this extensibility.

**Modules** Stratego programs are organized in modules, defined in separate files. A module can import other modules, making their contents accessible. Imports are transitive. Modules can define algebraic data type signatures, rewrite rules, and strategies.

**Strategic Rewriting** In Stratego, transformations are defined using term pattern matching (`?p`) and term pattern instantiation (`!p`) as basic operations (also known as match and build). For example, a rewrite `?Add(x, y); !Add(y, x)` swaps the subterms of an `Add` term, by sequentially composing a match and a build. Pattern matching is a first-class operation. That is, it is not bound to the use in a pattern matching construct that handles all cases. Thus, a pattern match fails when it is applied to a term that does not match, and is allowed to do so. Failure is first class (any transformation may

[1] One can argue with the choice of core language, but that is not the topic of this paper.

```
p  ::= b
b  ::= begin st* end
st ::= var x:t; | x:=e;
     | f(e1,...,en); | b
e  ::= x | f(e1,...,en)
t  ::= void
```

```
i ::= [0−9]+
e ::= i | e1+e2 | e1*e2
    | e1&e2 | e1|e2 | ...
t ::= int
```

```
st ::= if e then st* else st* end
     | if e then st* end
     | while e do st* end
     | for x := e1 to e2 do st* end
```

```
module desugar/core

desugar = innermost(Desugar)

Desugar = fail
```

```
module desugar/int
imports desugar/core

Desugar = BinOpToCall

BinOpToCall :
  f#([e1, e2]) → ⟦ f(e1, e2) ⟧
  where <is-bin-op> f

is-bin-op :
  ?"Add"
  <+ ?"Mul"
  // etc.
```

```
module desugar/control
imports desugar/core

Desugar =
  ForToWhile <+ IfThenToIfElse

ForToWhile :
  ⟦ for x := e1 to e2 do st* end ⟧ →
  ⟦ begin
      var x : int; var y : int;
      x := e1; y := e2;
      while x <= y do
        st* x := x + 1;
      end
    end ⟧
  where new ⇒ y

IfThenToIfElse :
  ⟦ if e then st* end ⟧ →
  ⟦ if e then st* else end ⟧
```

```
module eval/core

eval =
  eval-special
  <+ all(eval); try(eval-exp)

eval-special =
  EvalVar <+ eval-stats
  <+ eval-assign
  <+ eval-declaration

eval-assign =
  ⟦ x := <eval ⇒ e> ⟧
  ; rules( EvalVar.x : ⟦ x ⟧ → ⟦ e ⟧ )

eval-declaration =
  ?⟦ var x : t; ⟧
  ; rules( EvalVar+x :- ⟦ x ⟧ )

eval-stats =
  Stats({| EvalVar : map(eval) |})
eval-exp = fail
```

```
module eval/int
imports eval/core

eval-special = eval-or <+ eval-and

eval-exp =
  EvalAdd
  <+ EvalMul
  // etc.

EvalAdd: ⟦ Add(i, j) ⟧ → ⟦ k ⟧
    where <addS> (i, j) ⇒ k

EvalMul: ⟦ Mul(i, j) ⟧ → ⟦ k ⟧
    where <mulS> (i, j) ⇒ k
```

```
module eval/control
imports eval/core

eval-special =
  eval-if <+ eval-while
  ; eval

eval-if =
  ⟦ if <eval> then <*:id> else <*:id> end ⟧
  ; EvalIf; eval-stat

eval-while :
  st@⟦ while e do st* end ⟧ →
  ⟦ if e then st* st else end ⟧

EvalIf :
  ⟦ if i then st1* else st2* end ⟧ →
  ⟦ begin st1* end ⟧
  where <not-zero> i

EvalIf :
  ⟦ if 0 then st1* else st2* end ⟧ →
  ⟦ begin st2* end ⟧
```

**Figure 1** The TFA language. Columns are modules core, int, and control, rows are aspects grammar, desugaring, and evaluation.

fail) and is handled by the choice combinator **<+**. The choice s1 **<+** s2 between two transformations first applies s1 and when that fails applies s2.

In general, Stratego provides composition of transformation strategies from pattern matching and instantiation using a small set of combinators for control (including sequential composition and choice) and generic term traversal.

**Named Strategies and Rules** A strategy definition f = s names a strategy expression s and can be invoked as a combinator using its name. For example, the definition desugar = innermost(Desugar) defines the desugar transformation as the application of the innermost strategy with the Desugar rule(s). A rewrite rule f : p1 → p2 **where** s is sugar for a strategy f = {x,..: **?**p1; **where**(s); **!**p2}, i.e. it rewrites a term matching p1 to an instance of p2 provided that the condition s succeeds. Using these features, we can define many small transformation components (rules, strategies) and make multiple different compositions from these basic components. For example, we can compose rewrite rules using choice (e.g. EvalAdd **<+** EvalMul), to apply multiple rules to a term.

**Open Extensibility** The key to the extension of transformations illustrated in figure 1 is Stratego's open extensibility of rules and strategies. A module may provide several definitions for the same name. And such definitions may be defined across several modules, independently of each other. For example, the desugar/core module defines the desugar strategy in terms of the Desugar strategy. It defines the latter as the **fail** strategy, which always fails. Thus, applying the desugar strategy as defined in that module performs the identity transformation. However, the desugar/int and desugar/control modules (independently) extend the definition of Desugar. When combining these modules the definitions of Desugar are combined to

```
Desugar = fail <+ BinOpToCall <+ ForToWhile <+ IfThenToIfElse
```

and the desugar strategy normalizes a term with respect to those rewrite rules.

This extensibility feature is the core reason that Stratego has a whole program compiler. All modules of a program are combined in order to gather and combine all the extensions of each strategy. In addition, there are two more features that complicate separate/incremental compilation: dynamic rules and overlays.

**Dynamic Rules** Dynamic rules are rewrite rules that are defined dynamically during a transformation using the **rules**(...) construct. For example, the evaluation strategy in figure 1 uses a dynamic rule EvalVar to map variables to values. Furthermore, a dynamic rule gives rise to a number of derived strategies for applying and reflecting on the dynamic rule. There are currently 18 such derived strategies. Only one definition of each of these derived strategies should be generated per dynamic rule name, even if it is defined in multiple modules. Therefore, this feature requires global program information.

**Overlays** Overlays (which do not appear in the example in figure 1) are pattern aliases that can be used in both match and build position [23]. The applications of overlays are syntactically indistinguishable from regular terms. Overlays are greedily expanded; at runtime they are no longer available by name. A module does not need to import the module that defines the overlay for it to be available. Therefore the compiler needs to know all overlay definitions in the entire program to identify all their uses and expand them, affecting incremental compilation.

**Table 1** Code metrics of the WebDSL compiler code base.

| | |
|---|---|
| # of modules | 399 |
| # of distinct named strategies | 10 091 |
| # of congruences | 769 |
| # of dynamic rule helper strategies (18 × 371) | 6678 |
| # of user-defined strategies (rest) | 2644 |
| # of modules in which a named strategy is defined | 10 091 |
| 1 module | 9234 |
| 2–10 modules | 826 |
| 11–20 modules | 19 |
| 21–30 modules | 8 |
| > 31 modules | 4 |
| # of possibly ambiguous strategy use sites | 942 |
| # of overlay definitions | 19 |
| # of modules in which a distinct overlay is used | 19 |
| 0 modules | 8 |
| 1 module | 9 |
| 2 modules | 1 |
| 23 modules | 1 |
| # of distinct dynamic rule names | 371 |
| # of contributions per dynamic rule name | 371 |
| 1 contribution | 334 |
| 2 contributions | 26 |
| 3 contributions | 5 |
| 4 contributions | 3 |
| 8 contributions | 2 |
| 10 contributions | 1 |

**Open Extensibility in the WebDSL Compiler** The extensibility features discussed above are used in practice in compilers for languages used in production. For example, WebDSL, a domain-specific web programming language [8, 10], consists of a number of sublanguages such as for the definition of data models, querying, computation, user interface templates, and access control [9, 25]. The WebDSL compiler uses the open extensibility of Stratego to define extensions to basic analysis and transformation strategies per sublanguage. We present some relevant statistics of the WebDSL language definition in table 1 to show the relevance of the previously identified Stratego features in a real-world language definition.

The code base of the compiler consists of 399 modules with 2644 distinct strategies in the source code. Many additional strategies are generated as helpers of dynamic rules (6678), and constructors of terms (769) are usable as strategies as well. A total of 10091 strategies are compiled by the stratego compiler when the entire project is built. Out of those strategies, 857 have definitions in more than one module, showing how much the cross-modular extension feature is used.

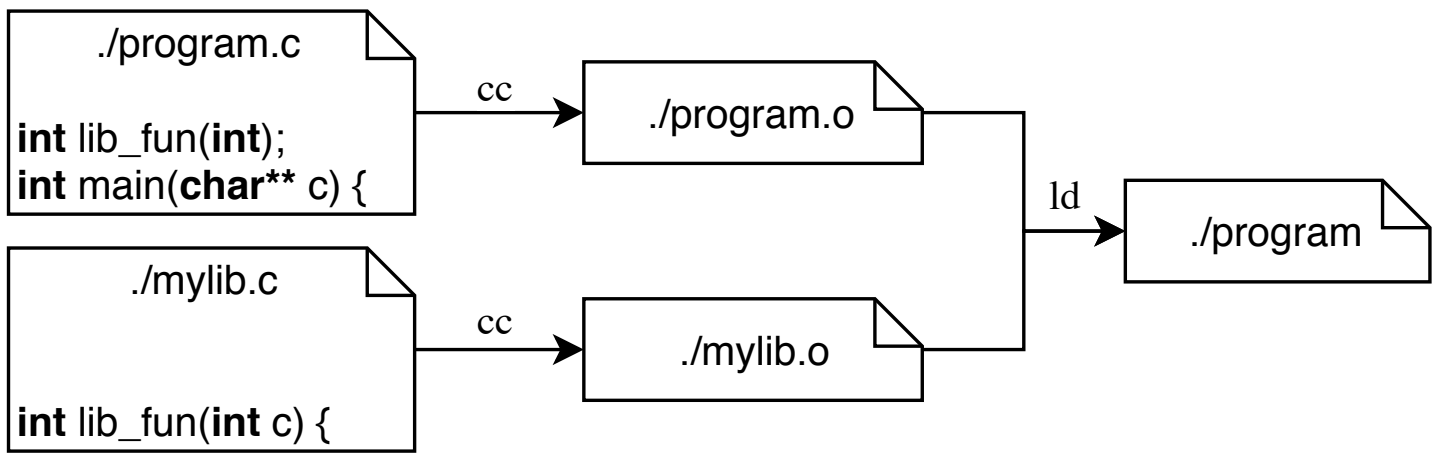

**Figure 2** An example build of a simple C project.

Ambiguous strategy use sites are numerous within the codebase, 942 calls to a strategy are initially ambiguous in arity. Overlays are not used very much, only 19 definitions exist in the codebase and 8 of those are not even used any more. Most overlays are only used in one module (usually the defining module), although one is used in 23 modules.

The project uses 371 dynamic rules, most of which (334) have only one definition. These are most likely used as scoped mutable variables, rather than proper dynamic rewrite rules. That still leaves 37 dynamic rules with multiple definitions in different places.

## 3 Compiler Architectures and Build Systems

Compilation of Stratego's open extensibility requires the integration of definitions from multiple modules, precluding a simple separate compilation model. In this section, we first recall how in separate compilation (for the C language) a compiler outsources dependency tracking and computing which files to recompile to a build system; an idea we will build on for our incremental compiler. Then we review the original whole program compilation model for Stratego and its limited support for separate compilation. We also discuss a previous attempt at incremental compilation by dynamic linking.

### 3.1 Separate Compilation and Build Systems

Separate compilation allows efficient recompilation of programs consisting of multiple compilation units. Only the compilation units that are affected by a change need to be recompiled. Typically, the work of determining which compilation units to recompile is not done by the compiler, but by an external build system. This is a nice separation of concerns. The compiler is not complicated by tracking dependencies and whether target files are up-to-date and the build system has to know only the external interface of the compiler.

For example, consider the C programming language and the compilation scenario illustrated in figure 2. C has a separate compiler, which translates a .c source file and to .o object file. Cross-module static checking is enabled by means of 'forward declarations' of functions (which are typically defined in additional header files included during preprocessing), defining the required interface to be implemented by the other files. Object files are combined into an executable by a linker. Thus, building a C program involves compiling all its files and then linking the resulting object files. Most C projects use a build system such as Make for this process such that only the files that are changed (or are affected by a change in a header file) need to be recompiled.

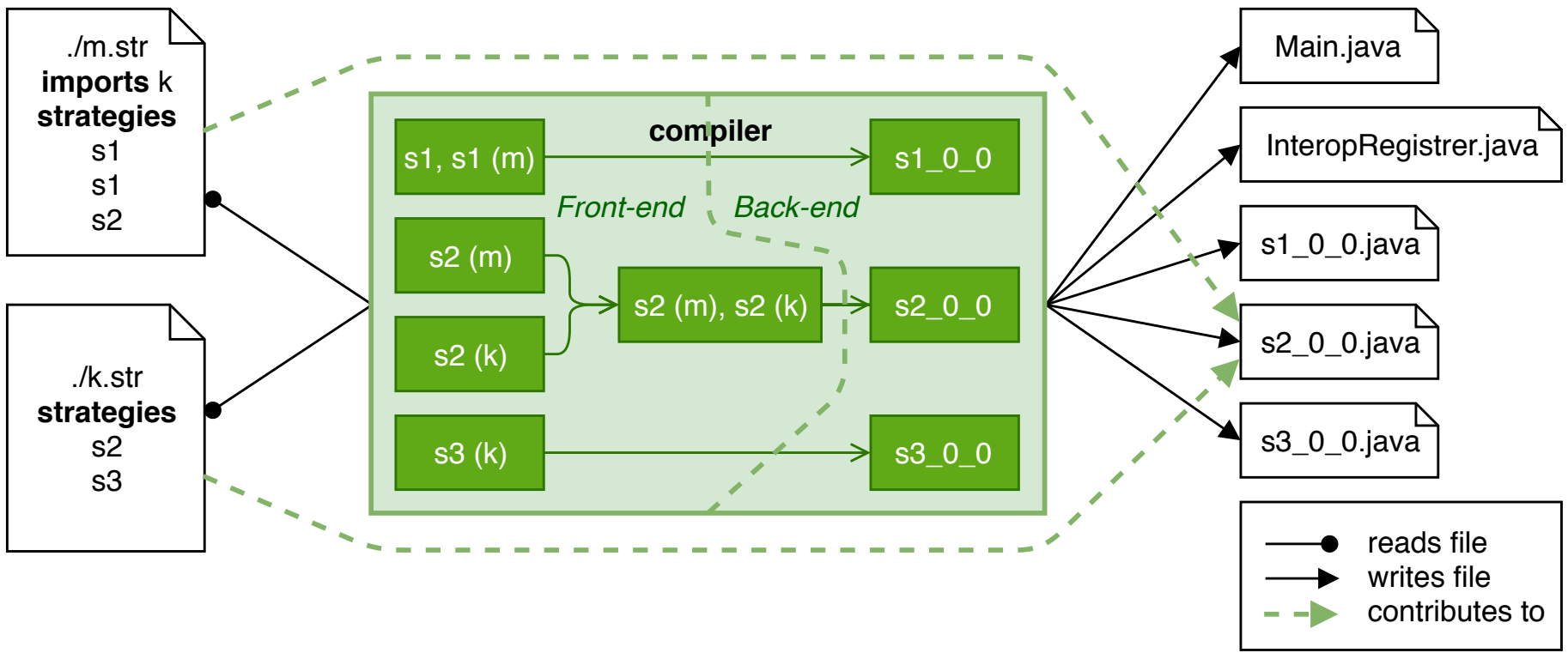

**Figure 3** Whole-program compilation example for Stratego. All modules are discovered and merged by a single process.

While this separation of concerns between compiler and build system is attractive for the compiler writer, it is less so for the programmer who needs to make sure the build system configuration accurately reflects the dependencies in the program often leading to unsound configurations, requiring frequent expensive from-scratch builds.

### 3.2 Whole-Program Compilation

The simple separate compilation model of C does not apply to Stratego because strategy definitions across modules need to be integrated. Therefore, Stratego's compiler uses a whole program compilation model [3] as illustrated in figure 3. (The original compiler translated Stratego to C. The current compiler produces Java code. The compiler architecture is unchanged.) The compiler reads all module files by following imports from the main file of a Stratego program. After parsing the files, the compiler builds a single internal model of the program. Strategy names (which can be overloaded in addition to have multiple definitions) are disambiguated by including the arity in the name. Then multiple definitions with the same name are merged into a single definition. Strategy definitions are merged by renaming the arguments to be consistent, scoping each strategy body, and defining the merged body as a choice between the two bodies, as illustrated in figure 4. The order between definitions from different modules is undefined. The back-end of the compiler translates the simplified, merged definitions to Java classes and extracts information such as term constructors, constants, and strategy names to put into the Main and InteropRegistrer classes.

```
desugar(|e) = CallT(\SVar(n) → n\, id , id)

desugar(|env): Var(e) → <lookup> (e, env)
```

⇒

```
desugar(|e1) =
  CallT(\SVar(n) → n\, id , id)
  <+ {e: (Var(e) → <lookup> (e, e1))}
```

**Figure 4** Integration of strategy definitions with the same name.

This process is illustrated by figure 3. Strategies s1 and s3 are only defined in one module, and each have their corresponding Java class in the compiler output. Strategy s2 is defined in both modules, and merged by the compiler into a single Java class.

**Limited Separate Compilation Support** While the Stratego compiler is a whole-program compiler, it has support for separate compilation of libraries. Compiled libraries consist of an interface (list of strategies and term constructors), and the Java classes with the compiled code. However, these libraries are second-class citizens. A compiled library can only be imported in its entirety since its internal module structure is discarded. Furthermore, separate compilation is achieved by limiting open extensibility to library boundaries; an (external) strategy defined in a library can not be redefined or extended. Dynamic rules defined in a library cannot be extended either.

To recover some of the expected functionality for external strategies, Stratego provides two special keywords, `extend` and `override`, which can be applied to a definition that is already defined in an imported library. The `override` keyword will override the definition with the local one, while the `extend` keyword overrides the definition but can call the original through the `proceed` keyword. These modifiers can only be applied to a single definition, i.e. these definitions cannot be further extended like regular strategy definitions. Calls to `proceed` have the same overhead as that of a normal strategy call, thus making this form of strategy extension more expensive at run-time too.

Thus, separate compilation allows the creation of libraries with reusable functionality, but without the idiomatic extensibility features of Stratego to interact with that code.

## 3.3 Independent Compilation with Dynamic Linking

In a case study of the Stratego compiler for the Pluto incremental build system, Erdweg, Lichter, and Weiel developed a "file-level incremental" compiler for Stratego [6]. To understand that model, we first discuss the Stratego to Java compilation scheme.

**Java Compilation Scheme** Figure 5 illustrates the compilation scheme. Each compiled strategy definition is a class that extends the `Strategy` abstract class. This class has invocation methods for a number of different arities with a default implementation that dispatches to a general method for two arrays of arguments (the strategy and term arguments). A generated strategy class extends this abstract class and implements the general method.

Strategy names are translated to the same name with underscores instead of dashes and followed by their arity separated by underscores. Each class has a public static

```
desugar(|env): Var(e) → <lookup> (e, env)
```

⇓

```
public class desugar_0_1 extends Strategy {
  public static Strategy instance = new desugar_0_1();

  public IStrategoTerm invokeDynamic(IStrategoTerm term, Stratego[] sargs, IStrategoTerm[] targs) {
    // ... match Var, bind e to local variable, build pair, bind to current term ...
    term = lookup_0_0.instance.invoke(term);
    return term;
  }
}
```

**Figure 5** Stratego to Java compilation scheme.

field instance where the instance of the strategy is saved. This field is used to be able to override or extend strategies in compiled libraries. The overriding or extending strategy is a new class that extends the original class, overrides the invocation method, and overrides the instance field of the original class with an instance of itself.

The Main class that the backend generates provides an entry point into a Stratego program. If a main strategy is defined, it can be called through the Main class, which will set up the execution context, compute shared constants and term constructor objects, and execute the program. The shared constants and term constructor objects are stored in fields of the Main class and referenced by the other classes.

The InteropRegister registers all compiled strategies in the context object for interoperability with the Java implementation of the Stratego interpreter. This hybrid interpreter can load compiled libraries like the Stratego standard library through these InteropRegisters, and then interpret the main program as an AST which was only compiled with the compiler frontend.

**Dynamic Linking Compilation Scheme** Erdweg, Lichter, and Weiel experimented with a separate compilation model for Stratego by relying on dynamic linking. To the best of our understanding from inspecting the code, this was an independent compiler, i.e. it skipped cross-module static analyses. The case study demonstrates the applicability of the Pluto build system to dependency tracking and rebuilding within a compiler. However, no speedup is reported on, nor were the changes merged into the Stratego compiler.

The compilation model is illustrated in figure 6. The compiler produces for each Stratego source file a list of Java classes for the strategy definitions in that file. With some changes to the Stratego runtime system, the Java classes are merged at run-time rather than merged statically. At the start of the program, all loaded classes register themselves by name to a StrategyCollector, as illustrated in figure 7.

The Main class initializes fields for every strategy by looking up the strategy by name from the collector. The collector builds StrategyExecutor objects for this which contain an array of Strategy classes that are produced by the compiler. These are attempted in order when the StrategyExecutor is invoked. Although the hashmap lookup based on

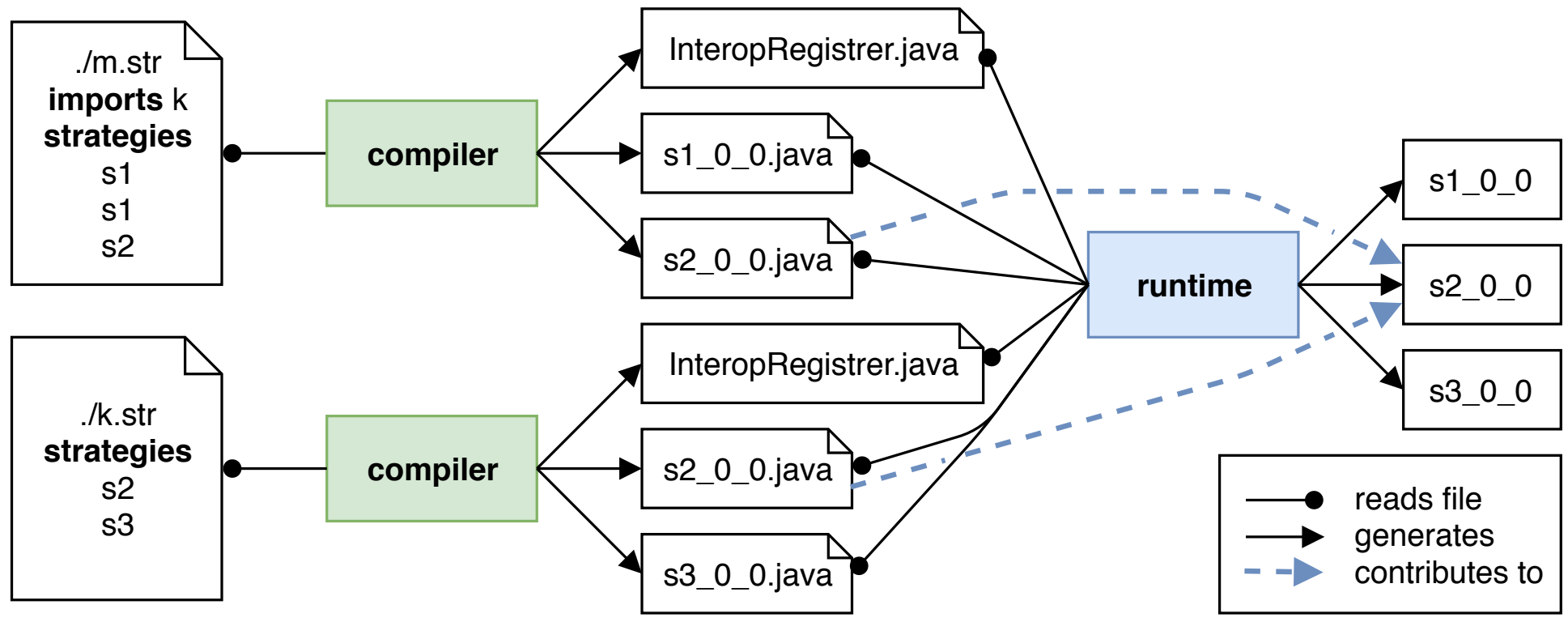

**Figure 6** Dynamic linking, incremental compilation example for Stratego. The compiler is run for each module, governed by the Pluto build system. An adapted runtime merges strategies dynamically, i.e. during program start-up.

```
desugar(|env): Var(e) → <lookup> (e, env)
```

⇓

```
public class desugar_0_1 extends RegisteringStrategy {
  public static final Strategy instance = new desugar_0_1();

  public static Strategy getStrategy(Context context) {
    return context.getStrategyCollector().getStrategyExecutor("desugar_0_1");
  }

  public void registerImplementators(StrategyCollector collector) {
    collector.registerStrategyImplementator("desugar_0_1", instance);
  }

  public void bindExecutors(StrategyCollector collector) { ... }

  public IStrategoTerm invokeDynamic(IStrategoTerm term, IStrategoTerm arg1) {
    // ... match Var, bind e to local variable, build pair, bind to current term ...
    term = Main.lookup_0_0.invoke(term); // note the difference in invocation!
    return term;
  }
}
```

**Figure 7** Compilation scheme with dynamic linking.

strategy name is only done once for every strategy call, each strategy definition from a different module requires a separate invocation in a different class.

### 3.4 Summary

Whole-program compilation requires minimal configuration and is therefore easy to use and requires less maintenance. However, whole-program compilation does not scale because recompilation time is proportional to the size of the entire program instead of the size of the change. We would like to keep the low-configuration benefit of whole-program compilation, while scaling incrementally with the size of the change. Pure separate compilation cannot be applied, as whole-program knowledge is required for compilation. The disadvantage of the dynamic linking model is that it shifts the burden of integrating strategy definitions to run-time, which incurs execution overhead and requires disruptive changes to the compilation scheme. Therefore, we need a hybrid-incremental compilation approach, which we will describe and apply to Stratego in the next section.

## 4 Incremental Compilation

In this section we first present our approach to incremental compilation in general. Then we make this concrete by presenting the application of our approach to the Stratego language and compiler.

### 4.1 A General Blueprint for Incremental Compilation

Our approach to incremental compilation results in a compiler that, from an external interface, looks like a whole-program compiler, not a separate compiler. Internally it uses an incremental build system that handles dynamically discovered dependencies. This build system caches intermediate results within the compiler to make it incremental.

**Data Splitting** The key idea is that every file that the compiler reads, is split up into parts as soon as possible. Each of these parts is then processed by separate build tasks. This allows the build system to pick up on parts that are unchanged in the file and parts that are changed. This idea is what gives the compiler sub-file incrementality.

The size of the parts influences how incremental the compiler is. A coarse-grained split results in less incrementality as more code unrelated to a change is recompiled. A fine-grained split increases the interaction between the parts during compilation steps, so that the overhead of the build system tracking all the dependencies becomes higher than the gains from avoiding repeated compilation of unchanged code.

The split to choose depends on the language features, and how much of the original compiler to reuse as is. A reasonable split is to choose parts that become separate outputs of the compiler. However, a finer-grained split is also possible, where parts are merged in a later stage of the compiler.

**Compiler Stage Splitting** Using an incremental build system to compose a compiler allows defining compiler stages as build tasks. The outputs of build tasks are cached and can be reused on recompilation. More tasks means more dependency tracking and thus more overhead. The number of tasks mostly depends on the granularity of the data split. However, even stages that operate in succession on a single part of the data can be split to improve performance. The main idea behind splitting up a successive operations is to avoid the later (expensive) operations in the chain, if an early operation results in the same output despite the change in input. That is, if an early operation gives the same output for different inputs, and the following operations are expensive, it may be worth the build system overhead to split the operations over multiple tasks. This allows the build system to cut off early when it observes that an intermediate result has not changed. The trade-off should be made by taking into account the overhead of the build system for such a change, the chance that this situation happens, and the expense of the later operations.

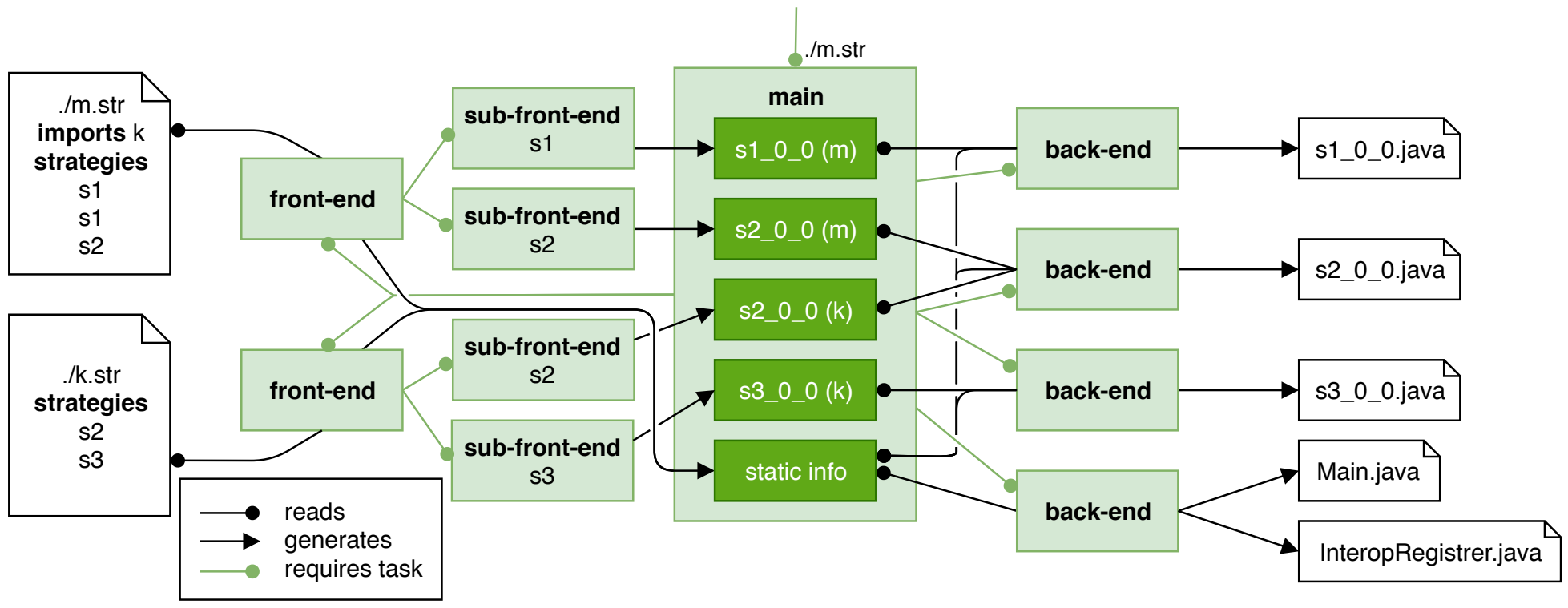

**Figure 8** Static linking, incremental compilation example for Stratego. A frontend task for each file, a backend task for each strategy.

### 4.2 Incremental Compilation for Stratego

To apply this blueprint to Stratego, we split modules by strategy definition and some extracted information for the static analyses at the end of the front-end of the compiler, as illustrated in figure 8. Then we merge strategy definitions with the same name and arity, and call the compiler back-end for each of these merged definitions. This results in one sub-front-end task per strategy in a file, one front-end task per file, and one back-end task per merged strategy, to minimise the work when a small change is made in a file.

The static analyses are done non-incrementally and were found to take an insignificant amount of time in our benchmarks (section 5). Other languages with more expensive static analyses may benefit from an incremental static analysis stage.

**PIE** As our build system we use PIE [14, 15], a sound and incremental build system supporting dynamically discovered dependencies to build tasks and files. Build tasks are regular (imperative) functions, except that they call other tasks (and can use their return value) by 'requiring' them, which instructs PIE to record a dependency. Similarly, build tasks can require (read from) or 'provide' (write to) files, which records dependencies to those files. PIE only re-executes tasks when their input changes, when the return value of a required task changes, or when a required/provided file changes. Otherwise, PIE returns the cached value of a task, providing incrementality implicitly, freeing the build developer from having to explicitly handle incrementality.

■ **Listing 1** The build orchestration algorithm, simplified to fit on one page. Tasks (which are incrementalized) are defined with the **task** keyword, and called (required) with **req**. The subFrontEnd, frontEndLib, and backEnd tasks are calls into parts of the original compiler, wrapped as tasks so they are incrementalized by the build system. The front-end tasks require the file corresponding to their input module, and the back-end tasks provide the generated files, ensuring re-execution when the input files change.

**(a)** Type definitions and combining front-end information into the static information struct.

```
type M = String // Module name
type S = String // Strategy name
type C = String // Constructor name
type AST = ATerm // Stratego Core AST
type SI = struct: // Static Information
  imps: Rel[M,M], // Imports
  defStrs: Rel[M,S], // Defined strategies
  defCons: Rel[M,C], // ... / constructors
  usedStrs: Rel[M,S], // Used strategies
  usedCons: Rel[M,C], // ... / constructors
  strUsedCons: Rel[S,C], // ... per strategy
  strASTs: Rel[S,AST], // Definitions of strategies
  olayASTs: Rel[C,AST], // ... / overlays
type FI = struct: // Front-end Information
  imps: Rel[M,M],
  defStrs: Set[S],
  defCons: Set[C],
  usedStrs: Set[S],
  usedCons: Set[C],
  strUsedCons: Rel[S,C],
  strASTs: Rel[S,AST],
  olayASTs: Rel[C,AST],
type LI = struct: // Library Information
  defStrs: Set[S],
  defCons: Set[C],

func combineInfo(si: SI, mod: M, fi: FI):
  si.imps[mod] := fi.imps ∪ defaultImps
  si.defStrs[mod] := fi.defStrs
  si.defCons[mod] := fi.defCons
  si.usedStrs[mod] := fi.usedStrs
  si.usedCons[mod] := fi.usedCons
  si.strUsedCons ∪= fi.strUsedCons
  // Strategies with the same name go together:
  si.strASTs ∪= fi.strASTs
  si.olayASTs ∪= fi.olayASTs

func combineInfoLib(si: SI, mod: M, li: LI):
  si.defStrs[mod] := li.defStrs
  si.defCons[mod] := li.defCons

task frontEnd(M) → FI
task subFrontEnd(S) → FI
task frontEndLib(M) → LI
task backEnd(S, Set[AST], Set[AST])
```

**(b)** Pseudocode for the build orchestration.

```
task main(mainMod: M):
  W: List[M] := [mainMod] // Worklist
  // Seen modules to prevent loops for cyclic modules
  seenMods: Set[M] := {mainMod}
  staticInfo: SI := SI() // Static analysis info
  // FRONTEND + COLLECT STATIC ANALYSIS INFO
  while W is not empty:
    mod := W.pop()
    if mod.isLibrary():
      li := req frontEndLib(mod) // cached task call
      combineInfoLib(staticInfo, mod, li)
    else:
      fi := req frontEnd(mod) // cached task call
      combineInfo(staticInfo, mod, fi)
      // Follow imports
      W.pushAll(staticInfo.imps[mod] \ seenMods)
      seenMods ∪= imps[mod]
  // STATIC ANALYSIS
  staticChecks(mainMod, staticInfo)
  // BACKEND
  for unique (_, str) in staticInfo.defStrs:
    olayASTs′ := {}
    for con in staticInfo.strUsedCons[str]:
      olayASTs′ ∪= staticInfo.olayASTs[con]
    // cached task call:
    req backEnd(str, staticInfo.strASTs[str], olayASTs′)


func staticChecks(main: M, si: SI):
  visStrs := si.defStrs; visCons := si.defCons
  for scc in topoSccs(main, si.imps):
    visCons′ := {}; visStrs′ := {}
    for mod in scc:
      visStrs′ ∪= visStrs[mod]
      visCons′ ∪= visCons[mod]
      // propagate information from earlier iteration
      for m in si.imps[mod]:
        visStrs′ ∪= visStrs[m]
        visCons′ ∪= visCons[n]

    for mod in scc:
      visStrs[mod] := visStrs′
      visCons[mod] := visCons′
      assert(usedStrs[mod] ⊆ visStrs′)
      assert(usedCons[mod] ⊆ visCons′)
```

**Build Algorithm** In listing 1 we present the build orchestration algorithm. It includes type definitions and helper functions in listing 1a, which combine information from one front-end task into the available global information. Note the commented line, which combines maps of strategy name to strategy definition AST. This is the point where strategies from different modules are combined by name.

Listing 1b contains the pseudocode of the build algorithm. The `main` task which orchestrates the build uses a worklist, starting from the main module, to run frontend tasks for each module. New modules are found through imports of processed modules.

The output of the front-end calls are collected and merged by strategy or constructor name, which is used in the static analysis (`staticChecks` function). Note that `staticChecks` is a function, not a task, and is therefore not cached, because it requires global information which changes almost every run, making caching moot. If `staticChecks` succeeds, the backend is called once per strategy, which takes a list of ASTs for that strategy (the contributions from different modules) and a list of ASTs for overlays that are used in the strategy.

**Static Analysis** In order to be compatible with the original whole program compiler, we reimplement the static analysis in a `staticCheck` function that takes in the gathered static information.

We use a topological order over the import graph, so we can easily propagate the strategy definitions according to the transitive import semantics. But since the graph can have cycles, we process the strongly connected components (SCCs) of the import graph instead of individual modules. In other words, a group of modules that import each other is processed together.

The visible strategies and constructors are computed first, then the used strategies and constructors are checked to be a subset for each module. To fit all the code on one page, we have elided some other static analyses from the pseudocode in listing 1. This includes some analysis information that is used by all back-ends. The following analyses are included in the implementation:

**Ambiguous Strategy Call Resolution** Ambiguous strategy calls result from the use of bare strategy names in strategy argument position. We resolve these names by looking for strategies of any arity with the same name. If a strategy exists with arity 0/0, the reference must resolve to this strategy and is considered unambiguous. When this arity does not exist, but there is only one arity available, the name can also be resolved. In other cases, where zero or more than one strategies are found, we report an error. The resolved names are returned by the static analysis. The resolutions are used in the back-end to replace the original names with the ones that include arity.

**Extending and Overriding Strategies** To be able to extend or override a strategy, an external strategy from a library needs to exist. External strategies are declared by a library along with the compiled implementations. We check for non-overlap of the relevant sets and give errors on any that are found.

**Cyclic Overlay Check** While looking into the original compiler details for overlays, we discovered a missing check. The original compiler will greedily expand overlays.

Therefore if an overlay definition **overlays** A() = A() will cause the compiler to loop. We added a check to fix this bug, which checks if there for cycles between any overlays (self-loops and indirect cycles). Overlays are only pattern aliases, without conditionals, so this analysis can be complete. We build a dependency graph of overlays and check for cycles.

**Implementation Effort** The incremental Stratego compiler was developed by one of the authors over a period of 10 months. Given the time spent on other tasks, this comes down to approximately 6-7 months of full-time equivalent. Before the start of the project the author was not very familiar with the compiler internals of Stratego.

The original Stratego compiler architecture certainly helped to make this quick development possible. The compiler generally transforms immutable trees (the AST), which makes it relatively easy to split off parts of it into cacheable tasks. The complicating factor is some mutable data structures that are used for extra (analysis) information. The mutation of these data structures would not be cached if naively separated into tasks to be cached.

**Conclusion** Our incremental compiler is an instantiation of our general blueprint for hybrid-incremental compilers. We use an incremental build system internally to incrementalize compilation tasks. Our compilation scheme splits files by top-level definition resulting in incrementality per top-level definition. It reuses the front-end of the original compiler on each top-level definition separately, and caches the results. After the front-end, the compiler performs all static analyses, merges top-level definitions with the same name and arity, and reuses the backend of the original compiler on each top-level definition separately. The output is completely backward compatible with the original runtime system.

## 5 Evaluation

We evaluate the performance of our incremental Stratego compiler by comparison against the original compiler. In particular, we evaluate the following research questions:

**RQ1** Does the incremental compiler scale with the size of the changes?

**RQ2** What is the overhead of a clean build with the incremental compiler?

**RQ3** Is the incremental compiler correct with respect to the original compiler?

### 5.1 Research Method

In a controlled setting, we compare the performance of the original compiler and our incremental compiler. We run our experiments on a MacBook Pro (Early 2013) with an Intel Core i7 2.8 GHz CPU, 16 GB 1600 MHz DDR3 RAM, and an SSD. The machine is running Mac OS 10.14.5, Java OpenJDK 1.8.0_212, Docker 2.0.0.3 (31259), VirtualBox 5.2.8 r121009 (Qt5.6.3), and Spoofax 2.5.8. As the experiments are performed inside a virtualised environment (Docker), the absolute numbers of our measurement may be higher than typical usage of the compiler in a normal setting. However, the virtual machine image can be easily reused by others to reproduce our experiments.

**Subject** As a benchmark subject we use the WebDSL language implementation. We use the version control commit history from 2016 to August 2019, consisting of 200 commits.

**Data Collection** We perform measurements by repeating the following steps. We run a clean build with the incremental compiler. Then we step to the next commit and pass the list of changed files to the compilation system. We do this for each commit. We measure the elapsed time for each successive compilation step, through Java's `System.nanoTime`.

We warm up the JVM on which the compiler runs by compiling a project and 3 successive commits, followed by a garbage collection. This was established by repeatedly running the benchmark on a cold JVM for many commits and noting the stabilisation of result times within 3 commits. We measure each project in a separate JVM invocation. During the JVM invocation we warm up once, and repeat the measurements 5 times. The results are shown as stacked bar-plots (with the arithmetic mean for each part) with whiskers to show the sample standard deviation (SD) in compilation time.

Similarly we run the original compiler for each project. This is only done to check that the compilation time of the original compiler does not vary much. We warm up the JVM as before with 1+3 compilations, and afterwards measure for each commit, and again go through the history 5 times. The results are shown with whiskers to show the full range (R) in compilation time.

### 5.2 Results

**Scaling** Of all 200 commits, 2 commits failed to build and were excluded from the results. We inspected these and determined that these failures were due to bugs that were fixed in the next commit. They failed to build with the original compiler as well as the incremental compiler.

Figure 9 shows the results of our benchmark for the WebDSL codebase. We found 78 commits that do not change any Stratego files and are excluded from the plot for that reason. The first commit, which is not compiled incrementally is also excluded. The remaining 119 commits are sorted in two different ways for comparison. The number of changed source files are shown in the top plot on the x-axis, the bottom plot shows the sum of sizes of (abstract syntax) trees passed to sub-front-end tasks.

The number of changed source files does not have a direct correspondence with the amount of work the incremental compiler needs to do. Although it is in direct relation with the effort for parsing, subsequent parts of the compiler deal only with changed top-level definition. The plot shows how the total incremental compile time does not scale directly with the number of changed files.

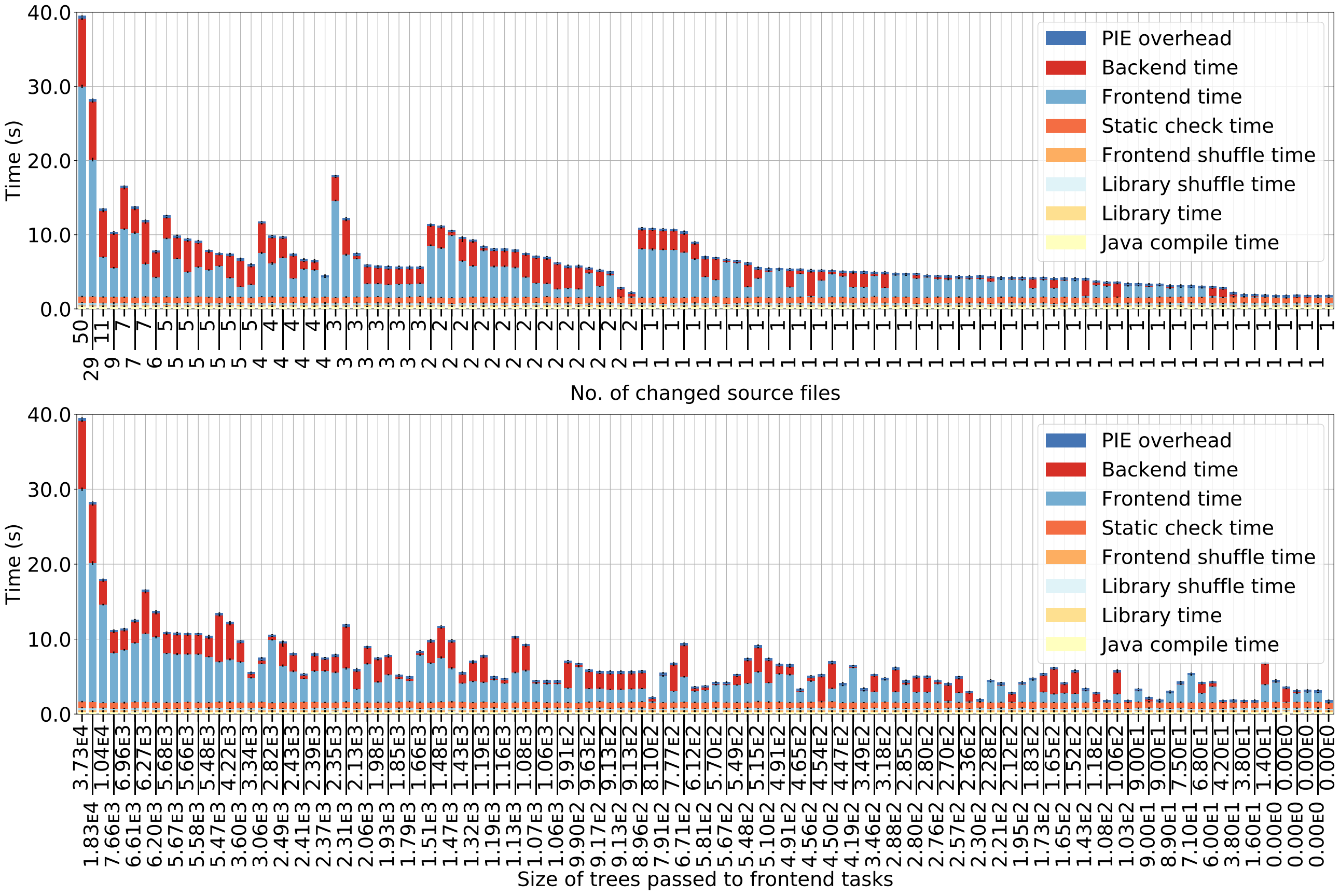

**Figure 9** Benchmark results of the incremental compiler on the WebDSL codebase from two perspectives. The results are displayed per commit, broken down into different parts of the incremental compiler. The Y-axis is time in seconds. The X-axis is number of changed files in the top plot, and the size of the trees passed to the frontend in the bottom plot.

The front-end times are closer, but still not one-to-one, correlated with the sizes of trees passed to them. Some Stratego language constructs are desugared into much larger pieces of code which are then processed further, so a small change to the text or tree can still have a non-proportional impact.

The code generation part of the compiler is a more straight-forward translation. If we would order the plot by sizes of trees passed to the back-end tasks, we would see a strict decline of the back-end times over the plot.

**Overhead** We compare the overhead of the incremental compiler during a clean build, as this is where the user of the incremental compiler will notice the overhead. For the WebDSL codebase, the observed clean build time for the incremental compiler is 178.53 seconds, whereas the build time of the original compiler is on average 93.54 seconds (see also figure 10). In other words, the incremental compiler has an overhead of 90.8 % for clean builds on the WebDSL codebase (RQ2). All incremental compilations are under the lowest fastest time with the original compiler. The overhead of the incremental compiler may have a variety of different causes.

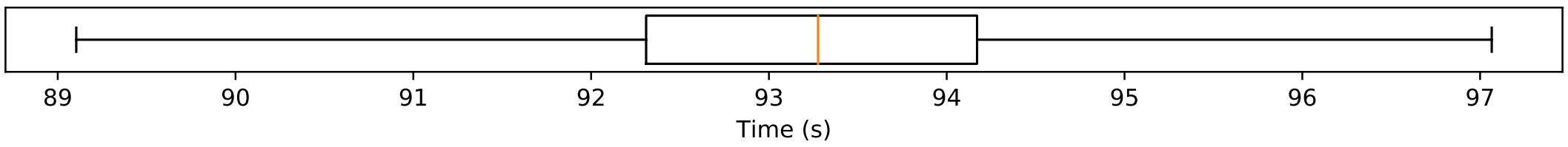

**Figure 10** Benchmark results of the original compiler on the WebDSL codebase. The X-axis is time in seconds. This is a boxplot over the measurements for all commits, as they have low variance. The whiskers span the full range of measurements.

**Correctness** To test the correctness of our compiler, we ran the full WebDSL compiler test suite. We used the result of compiling the WebDSL project with the incremental compiler and the original compiler. The two resulting WebDSL compilers had the same behaviour for all tests (RQ3) (in both successes and a number of failures, which were expected behaviour according to the WebDSL developers).

### 5.3 Interpretation and Discussion

Although the scaling behaviour of the incremental compiler does not entirely follow the size of changes to files in the benchmark, the general tendency is there (RQ1). When we inspect particular pairs of commits with unexpected ordering we find that unexpectedly cheap (in incremental compile time) commits change a strategy with the same name over multiple files. Unexpectedly expensive commits make changes in strategies that use expensive language features that are expanded by the compiler to much larger pieces of code.

Something we should note is a known issue with the official Stratego parser we use. This parser can take a relatively long time to parse certain Stratego files. This issue is solved in a new version of the parser, but that version is currently not stable enough for us to use in the benchmark.

The size of the overhead seems to reside in something other than the overhead of PIE itself and the shuffling of information. Our reimplementation of the static analysis of Stratego is also not the problem. It is unclear what exactly causes the overhead, but somehow the multitude of calls into the original compiler code or the separate treatment of the different files is causing overhead in the incremental compiler. We have attempted experimental changes to the compiler to verify or falsify assumptions on where this overhead comes from, but so far without success.

Compared to the time it takes for every compilation with the original compiler, a single clean build is a cost that anecdotally Stratego users are willing to pay for the speedup they receive on later incremental compilations. Despite the overhead that is visible in the clean build, our incremental compiler provides a speedup of up to $\sim 90\times$ for small changes in one or two files. Most commits take less than 10 seconds to compile, versus the 1.5 minutes that the original compiler takes, massively increasing productivity for Stratego programmers.

### 5.4 Threats to Validity

We explicitly discuss the threats to validity of this benchmark. In particular, we address generalizability of the results (external validity), factors that allow for alternative explanations (internal validity), and suitability of metrics for the evaluations goals (construct validity).

**External Validity** A threat to the generalizability of our results is that we only evaluated our Stratego compiler on Stratego code that originated from our research group. Therefore it has a certain code style that may influence the efficacy of our compilation approach. With the inclusion of WebDSL we cover all Stratego features and the classic Stratego code style with highly overloaded strategies, which are more expensive to compile incrementally.

**Internal Validity** A factor that allows an alternative explanation for our results is miscompilation, where only part of the program was actually successfully compiled, and another part was ignored. Early in the development of the compiler we found strange measurement results that came from the error recovery in the parser. Because of a misconfiguration, the Stratego parser sometimes failed to parse the entire file. Error recovery in the parser ensured that an abstract syntax tree was still returned, containing only those top-level strategy definitions that could be parsed successfully. Because our new compiler did not check if the parser had recovered from errors, parts of the Stratego program were not actually compiled. We found and fixed this problem through careful inspection of the benchmark results, and avoided more problems by running the WebDSL compiler test suite.

**Construct Validity** Regarding the suitability of metrics for the evaluation goal, we measure performance using elapsed time only. We control JIT compilation with a warmup phase. By running the garbage collector in between measurements, and monitoring the available memory, we control memory availability during all measurements. However, the incremental compiler stores intermediate results in memory and may perform differently in environments with less memory available.

Of course we can also not entirely eliminate background noise. However, we have repeated all measurements five times and see low variance between measurements (maximum sample standard deviation was 0.36 seconds). This is also visible in figure 9, where we display black whiskers on each stacked bar, all of which are barely distinguishable.

The virtualisation of our benchmark allows easy distribution of the benchmark for reproduction of our results. Since we use a long-running macro-benchmark, the virtualisation does not significantly influence the results. The same benchmark, run on the same hardware without virtualisation, shows similar results. There is only a 0 % to 25 % reduction in times, where the longest times were reduced most, which shows the overhead of virtualisation.

## 6 Related Work

**Recompilation** Independent compilation was an early way to speed up compilation by splitting up work and caching the intermediate results [1]. To reinstate guarantees of static analysis on the entire program, Mesa introduced separate compilation [7] and inspired other languages such as Modula-2 [28] and (an extension of) Pascal [12] to do the same. Incremental compilation was another refinement of the concept to allow a more arbitrary splitting of the work instead of by file [17, 18].

The novelty of our approach is to apply these (already over 30-year-old) ideas in a new way. We use an incremental build system to piece together parts of an existing non-incremental compiler into an incremental compiler. The expressive power of the build system we use, with dynamically discovered dependencies, allows us to fully express the build of the language inside the compiler. Therefore, the build system becomes an invisible part of our compiler. The user does not need to configure it on a per project basis.

**Build Systems** Relevant incremental build systems are of course the system we use, PIE [15], as well as its predecessor Pluto [6]. Pluto was in some ways more powerful than PIE as it can handle circular build dependencies which require a registered handler to compute a fixpoint. PIE, however, has much less overhead because it does not support this feature, and because it can do so-called bottom-up builds [14]. A bottom-up build requires the list of changed files and directories and uses that and the previous execution of the build script to do the minimal updates necessary to have a consistent output again. This gives less overhead from the build system because parts of the dependency graph do not need to be traversed at all.

Of course there are other build systems, both those that support incrementality and those in which incrementality can be hacked in by dynamically generating new build scripts and calling into those. Mokhov, Mitchell, and Jones analyse the features of different build systems and give an overview of desirable features [16].

Our approach is that the build system is an invisible part of our compiler. The build is defined once and for all inside the compiler, therefore the user does not need to configure it on a per project basis.

**Incremental Compilers** While our compilation approach works well for reusing an existing compiler, there are other approaches to incremental compilation, especially when one is built anew.

An example of a different incrementalization approach is JastAdd, a Java compiler built entirely on Reference Attribute Grammars (RAGs). These RAGs gave support for incremental evaluation [19]. To do something similar for the Stratego compiler we would need to build a completely new Stratego compiler in terms of RAGs, which is a non-trivial amount of work.

Another example of an incremental compiler is rustc, the compiler for the Rust programming language. The rustc compiler has an experimental incremental compilation mode that caches intermediate results, automatically records (dynamic) dependencies between these results, and reuses cached results when a dependency chain has not been invalidated [29]. However, this incremental compiler mode is not enabled by default (at the moment of writing) [30], since it sometimes makes compilation slower because of the overhead of incremental computation. The difference with our approach is that the Rust compiler automatically tracks dependencies based on reads and writes of data structures, whereas in our approach, we explicitly state the units of computation and dependencies between these computations. While explicitly stating this requires more effort, the advantage is that we can tune the granularity of the incremental compiler, to prevent the case where fine-grained dependencies cause too much overhead.

**Extension Unification** Extensibility has been studied from different perspectives. Erdweg, Giarrusso, and Rendel created a classification of extensibility, where Stratego falls within 'extension unification' [5, §4]. Extension is the terminology for supporting "language extension of a base language [where] the implementation of the base language can be reused unchanged to implement the language extension" [5, §2.1, Def. 1]. Unification is the terminology for supporting "language unification of two languages [where] the implementation of both languages can be reused unchanged by adding glue code only" [5, §2.2, Def. 2]. Stratego has a composition of extension and unification in that languages can be extended from a base (pre-defined strategies), and the different extensions (contributions to a pre-defined strategy) can be unified unchanged.

**Expression Problem** The expression problem defined by Wadler has more stringent requirements [26]. There is a reason its name includes 'problem'. Over the years there have been many suggested solutions [13, 20, 27] and extensions of the problem statement [11, 31]. The cited papers are only a few examples.

Stratego does not have the strong static type system required by Wadler for the original expression problem. This gives the language some more flexibility to attain the other properties more easily. Both types and functions on those types can be extended, without a change to the original code. And as presented in this paper, Stratego can now be separately compiled. There is also no linear order in language extensions defined in Stratego, as required by Zenger and Odersky [31], nor is there any glue code for combining different language extensions as required by Kaminski, Kramer, Carlson, and Van Wyk [11].

**Language Definitions** From a different perspective, if we look at competitors in language definitions we might look at attribute grammar systems such as Silver [21] and JastAdd [4]. Silver has separate compilation and generates multiple separate Java classes so it can leverage the Java compiler for hybrid incremental compilation, if composed by an external build system. It also leverages some dynamic linking to allow more concerns to be compiled separately, and therefore make compilation more incremental [22]. We are not aware of a publication on JastAdd's compilation scheme.

## 7 Conclusion

In this paper we have presented a design approach for hybrid incremental compilers. These compilers look like a whole-program compiler from the outside, but leverage an internal, incremental build system for incremental compilation. The blueprint of this design includes considerations on how to split up data and how to split up compiler stages inside the build system. We have given a concrete example of the blueprint with an incremental compiler for Stratego.

To motivate the decisions we made in our instantiation of the blueprint, we have presented an analysis of the Stratego language and shown how different features require global information to be compiled. The particularly problematic feature is that of unifying top-level definitions with the same name into a single definition that attempts the different alternatives.

Our instantiation of the blueprint splits up a file into top-level definitions early, processes each, then combines all equal-named top-level definitions before moving on to code generation. The static analysis is reimplemented but all other parts of the compilation are recycled from the original compiler.

By going through the version control history of a large Stratego project, we have demonstrated the incrementality of the new compiler. The original compiler takes about $93 \pm 3$ seconds to compile the project. Our results show that all but one commit with 1 file changed are compiled in under 10 seconds, a majority of 1 file changed commits compile in under 5 seconds, and even the largest commit which changes 50 files is incrementally compilable in under 40 seconds.

We would like to see another real-world programming language gain an incremental compiler with our design approach. The internally used build system and reuse of compiler components should make it a reasonably sized project that can greatly benefit the users of the programming language. It would also strengthen our hypothesis that this approach is reusable for other programming language compilers.

**Acknowledgements** We would like to thanks the anonymous reviewers for their valuable comments and suggestions.

This research was supported by a gift from the Oracle Corporation and by the NWO VICI Language Designer's Workbench project (639.023.206).

## About the authors

**Jeff Smits** is a PhD candidate at Delft University of Technology, where he works on domain-specific languages, language workbenches, data-flow analysis, and compilers. His current work includes designing a DSL for declarative data-flow analysis specification, and separate compilation for a term-rewriting language. You can contact him at j.smits-1@tudelft.nl and find further information at https://www.jeffsmits.net/.

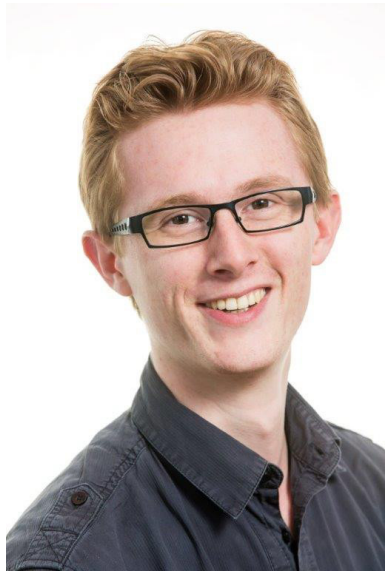

**Gabriël Konat** is a postdoctoral researcher at Delft University of Technology, where he works on domain-specific languages, language workbenches, bootstrapping, and incrementalization. His current work includes designing a DSL for programming incremental pipelines, improving the scalability of incremental pipeline algorithms, and applying incremental pipelines to language workbenches. You can contact him at g.d.p.konat@tudelft.nl and find further information at https://gkonat.github.io/.

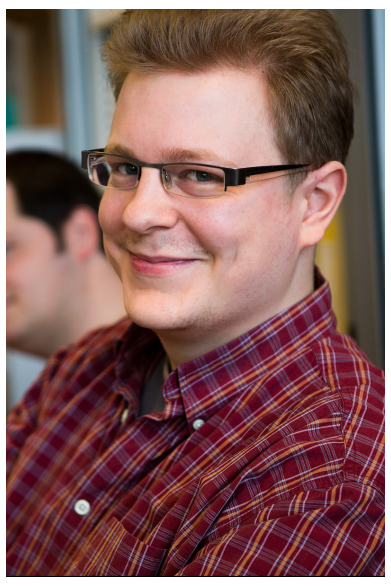

**Eelco Visser** is Antoni van Leeuwenhoek Professor of Computer Science and chair of the Programming Languages Group at Delft University of Technology. His current research is on the foundation and implementation of declarative specification of programming languages. You can contact him at e.visser@tudelft.nl and find further information at https://eelcovisser.org.

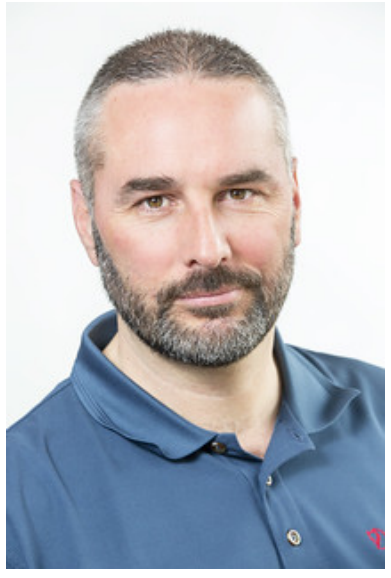